\documentclass{emulateapj} 
\usepackage{graphicx}
\usepackage{color,calc}
\usepackage{amsmath}
\usepackage{amssymb}
\usepackage{amstext}

\begin{document}

\title{The Dynamical Effects of White Dwarf Birth Kicks in Globular Star Clusters}
\shorttitle{The Dynamical Effects of White Dwarf Birth Kicks in Globular Star Clusters}
\submitted{Accepted for publication in ApJL} 
\author{John M. Fregeau\altaffilmark{1,2,3}, Harvey B. Richer\altaffilmark{4},
  Frederic A. Rasio\altaffilmark{5} \& Jarrod R. Hurley\altaffilmark{6}}
\shortauthors{FREGEAU, ET AL.}
\altaffiltext{1}{Kavli Institute for Theoretical Physics, UCSB, Santa Barbara, CA 93106, USA}
\altaffiltext{2}{fregeau@kitp.ucsb.edu}
\altaffiltext{3}{Chandra Fellow}
\altaffiltext{4}{Department of Physics and Astronomy, University of British Columbia, Vancouver, BC V6T 1Z1, Canada}
\altaffiltext{5}{Department of Physics and Astronomy, Northwestern University, Evanston, IL 60208, USA}
\altaffiltext{6}{Centre for Astrophysics and Supercomputing, Swinburne University of Technology, Hawthorn, Victoria 3122, Australia}

\begin{abstract}
Recent observations of the white dwarf (WD) populations in the Galactic globular cluster NGC 6397 suggest that 
WDs receive a kick of a few ${\rm km} \, {\rm s}^{-1}$ shortly before they are born.  Using our Monte
Carlo cluster evolution code, which includes accurate treatments of all relevant physical processes operating
in globular clusters, we study the effects of the kicks on their host cluster and on
the WD population itself.  We find that in clusters whose velocity
dispersion is comparable to the kick speed, WD kicks are a significant
energy source for the cluster, prolonging the initial cluster core contraction phase
significantly so that at late times the cluster core to half-mass radius ratio is a factor of 
up to $\sim 10$ larger than in the no-kick case.  WD kicks thus represent
a possible resolution of the large discrepancy between observed and theoretically
predicted values of this key structural parameter.  Our modeling also reproduces the observed trend
for younger WDs to be more extended in their radial distribution in the cluster
than older WDs.
\end{abstract}


\section{White Dwarf Birth Kicks}\label{sec:wdkicks}
Recent {\em HST} observations of the Galactic globular cluster NGC 6397 reveal
that its youngest white dwarfs (WDs) are significantly more extended
in their radial distribution within the cluster than their older visible counterparts
\citep{2008MNRAS.383L..20D}.  
Since the progenitors of the WDs are some of the
most massive stars in the cluster at $\sim 0.8\, M_\sun$, and since the older WDs
(of mass $\sim 0.5\, M_\sun$) have had time to diffuse to larger radii than where they formed
due to mass segregation, one would naively expect the opposite---namely that
the youngest WDs should be more centrally concentrated than their older counterparts.
One possible interpretation is that WDs receive a kick of a few ${\rm km} \, {\rm s}^{-1}$ shortly before they
are born \citep{2008MNRAS.383L..20D,2007MNRAS.381L..70H,2008MNRAS.385..231H,2008MNRAS.390..622H}.
Indeed, in open star clusters, where the velocity dispersion is $\sim 1\, {\rm km} \, {\rm s}^{-1}$
and the putative WD kick would be enough to eject it from the cluster, 
there is an observed relative dearth of WDs 
\citep{2003ApJ...595L..53F,2001AJ....122.3239K,1977A&A....59..411W}.

The mechanism generating WD ``natal'' kicks is likely asymmetric
mass loss late in the asymptotic giant branch (AGB) phase.  Although
AGB winds have not been directly observed, the effects of the resulting
mass loss have been, and there are good theoretical
reasons to expect such winds exist \citep{1993ApJ...413..641V}.
Alternatively, the kick could be produced as a result of small asymmetry
during the helium core flash (Ivanova, private communication).  Independent of precisely when in the evolution
of a star the kick occurs, the observed rotation rates of WDs
are consistent with non-axisymmetric mass loss at some point
in their evolution \citep{1998A&A...333..603S}.

Analogous to the hydrogen-burning main sequence in stars, star clusters
are thought to eventually enter a long lived binary-burning phase in
which inelastic dynamical scattering interactions of binaries with
other stars generate enough energy to prevent the cluster core from collapsing
\citep[e.g.,][]{2008gady.book.....B}.  The physics of the binary-burning
phase has been studied intensively.  It is only recently, however,
that different numerical methods have begun to agree on the structural
properties of a cluster in the binary burning phase 
\citep[at least in the equal-mass case;][]{2006MNRAS.368..677H,2007ApJ...658.1047F}.
Unfortunately, the predicted value of the core to half-mass radius ratio $r_c/r_h$
in the binary-burning phase is at least $\sim 10$ times smaller than what is observed for the majority
of Galactic globular clusters \citep[e.g.,][]{2007ApJ...658.1047F}.  
Although differences in the observational and theoretical
definitions of the core radius may account for part of the difference \citep{2007MNRAS.379...93H}, 
there is still a significant discrepancy.  If the bulk of Galactic clusters are indeed currently
in the binary-burning phase, a core radius larger than
expected from binary burning suggests the existence of an alternate core energy source.  
In this vein, it has been suggested that perhaps {\em tens} of Galactic clusters contain a
central intermediate-mass black hole \citep[IMBH;][]{2006astro.ph.12040T}, powering their cores
to the observed sizes.  Or the required energy could come from expedited
stellar evolution due to stellar collisions \citep{chatterjeeposter}, 
or prolonged mass segregation of compact objects \citep{2004ApJ...608L..25M,2007MNRAS.379L..40M}.
Alternatively, it could be that most clusters are simply not currently in the binary burning phase,
in which case their initial properties may play an important role \citep{2008ApJ...673L..25F}.

Since for a standard \citet{1993MNRAS.262..545K} initial mass function, roughly 15\%
of stars in a cluster will
become WDs within 12 Gyr, a WD birth kick of a few ${\rm km} \, {\rm s}^{-1}$ represents a potentially significant
core energy source (roughly 15\% percent of the total cluster energy for a WD kick speed comparable
to the velocity dispersion)---one that could resolve the discrepancy between theory and observations
on the structural parameters of clusters.  In this paper we consider the dynamical
effects of WD birth kicks by simulating the evolution of star clusters in which 
WDs receive kicks of a few ${\rm km} \, {\rm s}^{-1}$ at birth.  Our simulation method treats all relevant
physics in clusters self-consistently, as described below.  We find that WD birth kicks
can significantly prolong the initial cluster evolutionary phase of core contraction,
resulting in core sizes at late times that are larger than for clusters without kicks.
We also find that birth kicks produce an observational signature that is consistent with the observed
radial distributions of WDs in NGC 6397.

\section{Simulations}\label{sec:sims}

To simulate the evolution of clusters with WDs that experience birth 
kicks, we use the well-tested H\'enon Monte Carlo method, as described in detail
elsewhere \citep{2000ApJ...540..969J,2001ApJ...550..691J,2003ApJ...593..772F,2007ApJ...658.1047F,fregeaurasio2009}.
Briefly, it treats the effects of two-body relaxation in the orbit-averaged approximation
and, since it provides for a physical realization of the cluster at each timestep, has been extended to
include a treatment of all other relevant physics, including strong dynamical interactions of binaries,
physical stellar collisions, stellar evolution of single stars and binaries, and the effects of a 
Galactic tide.  It has shown excellent agreement with direct $N$-body methods for the evolution
of clusters with dynamically significant populations of binaries \citep{2007ApJ...658.1047F},
and those in which the effects of stellar evolution are dominant \citep{fregeaurasio2009}.
To model the effects of WD birth kicks we simply give a randomly oriented velocity kick to
a star as soon as it becomes a WD.  For simplicity, the kick is set to a fixed value between
2 and 9 ${\rm km} \, {\rm s}^{-1}$ for each simulation.  

Modeling individual clusters with any degree
of detail is a difficult endeavor \citep[e.g.,][]{2008AJ....135.2129H,2009arXiv0901.1085G}.  While we do not
attempt to model NGC 6397 in this paper, we find that our models do have roughly the same core
velocity dispersion at late times, and the model with $4 \, {\rm km} \, {\rm s}^{-1}$ WD kicks has
roughly the same ratio of core to half-mass radius at very late times 
\citep[to within a factor of 2 of the value in][]{2006AJ....132..447N}.
Our standard initial model is a $W_0=7.5$ King model with a virial 
radius of $5 \, {\rm pc}$, consisting of $N=3 \times 10^5$ objects, 1\% of which are 
binaries \citep[similar to NGC 6397's inferred initial binary fraction;][]{2008AJ....135.2155D}.  No 
initial mass segregation is assumed.  Single stars and binary primaries
are chosen from a \citet{1993MNRAS.262..545K} distribution from $0.15$ to $18.5\, M_\sun$.
Binary secondaries are chosen from a flat mass ratio distribution with a minimum at $0.15\, M_\sun$,
semimajor axes are chosen uniformly in the log from $5(R_1+R_2)$ (where $R_i$ are the stellar radii)
to the hard--soft boundary, and eccentricities are chosen from a thermal distribution truncated
at the upper end to prevent pericenter approaches smaller than $5(R_1+R_2)$.
Cluster metallicity is set to $Z=0.001$, as in \citet{2008AJ....135.2129H}.  
Although the metallicity of NGC 6397 is a factor of $\sim 5$ smaller, as described 
in \citet{2008AJ....135.2129H} stellar
evolutionary processes don't depend strongly on metallicity below $Z \approx 0.001$.
The Galactic tide represents that of a point-mass host galaxy of mass $10^{11}\, M_\sun$,
with the cluster in a circular orbit at $6.9\, {\rm kpc}$.

\begin{figure}
  \begin{center}
    \includegraphics[width=0.95\columnwidth]{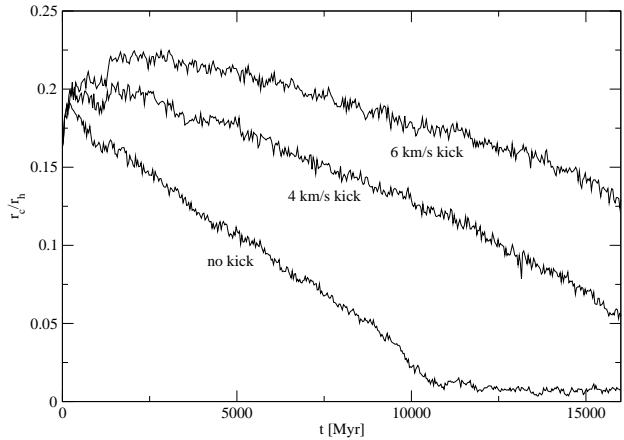}
    \caption{Evolution of core to half-mass radius ratio, $r_c/r_h$, for models with no WD birth kicks, 
      $4\,{\rm km} \, {\rm s}^{-1}$ birth kicks, and $6\,{\rm km} \, {\rm s}^{-1}$ birth kicks.
      \label{fig:rcrh}}
  \end{center}
\end{figure}

Figure~\ref{fig:rcrh} shows the evolution of $r_c/r_h$ for our standard model with no WD birth kicks, 
$4\,{\rm km} \, {\rm s}^{-1}$ birth kicks, and $6\,{\rm km} \, {\rm s}^{-1}$ birth kicks.
Around $11\,{\rm Gyr}$, the model without kicks enters the binary burning
phase, during which time $r_c/r_h\approx 0.01$.  \citep[Note that we use the standard three-dimensional, $N$-body definition of
the core radius;][]{fregeaurasio2009}.  The models with kicks show no evidence of reaching
a binary burning phase in their 16 Gyr of evolution.  Instead, they show a prolonged, gradual
contraction of the core with time, similar to the behavior found in $N$-body simulations 
by \citet{2007MNRAS.379...93H}. 
As expected, the model with $6\,{\rm km} \, {\rm s}^{-1}$ birth kicks displays an appreciably
larger core radius over time compared to the $4\,{\rm km} \, {\rm s}^{-1}$ kick model.
At late times, the models with kicks display large values of $r_c/r_h$, with values 
up to $\sim 10$ times larger than the model without kicks.  We note that for the model
with $4\,{\rm km} \, {\rm s}^{-1}$ kicks we find a value of $r_c/r_h \approx 0.11$ at
an age of 12 Gyr, which is significantly larger than the observed value of $r_c/r_h \approx 0.03$
for NGC 6397 \citep{2006AJ....132..447N}.
The $4\,{\rm km} \, {\rm s}^{-1}$ kick model reaches values of $r_c/r_h \approx 0.05$ at unphysically
late times, but further simulations are required to determine if a better match to NGC 6397 can be made
with a suitable choice of initial conditions.

\begin{figure}
  \begin{center}
    \includegraphics[width=0.95\columnwidth]{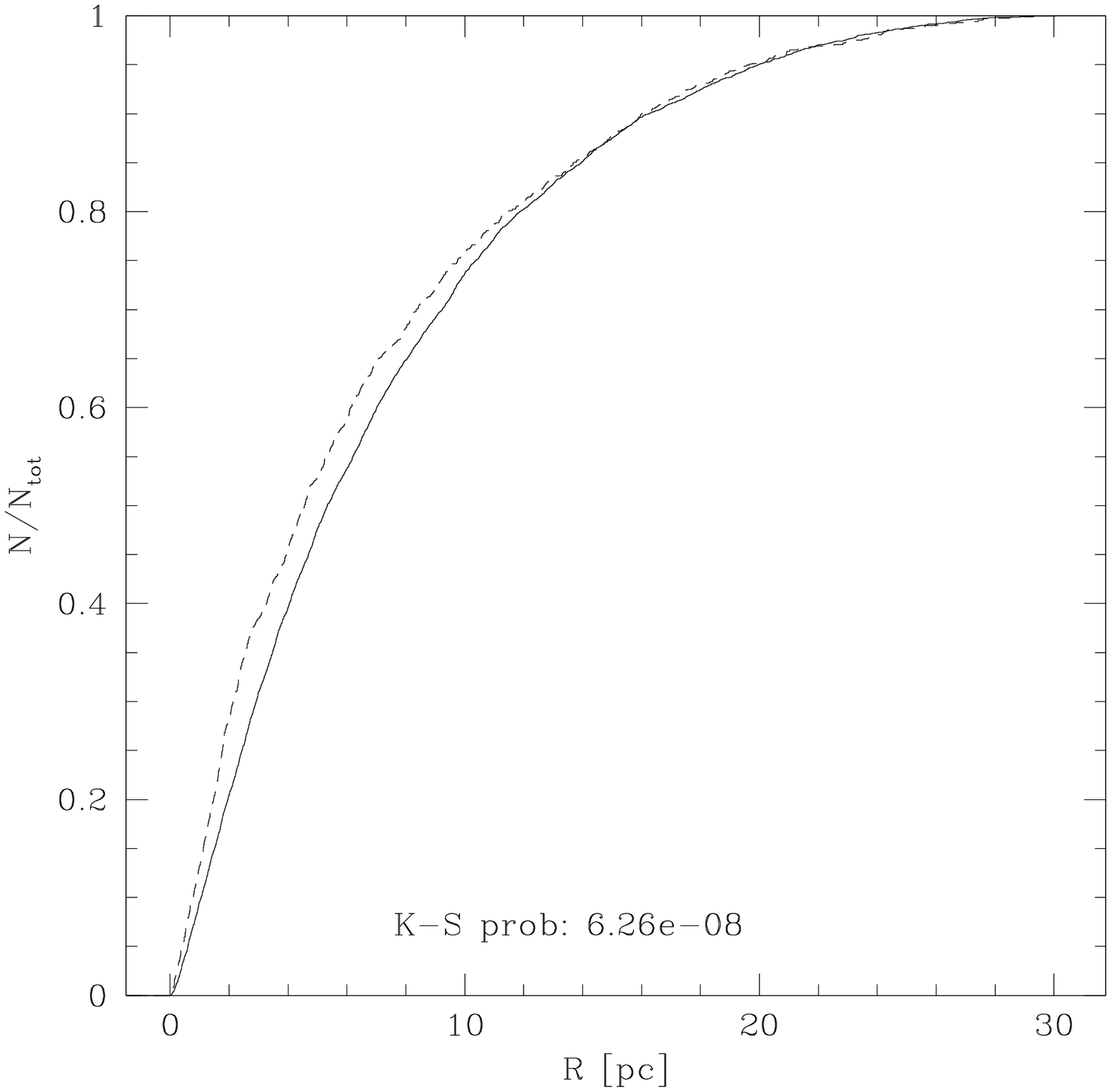}
    \includegraphics[width=0.95\columnwidth]{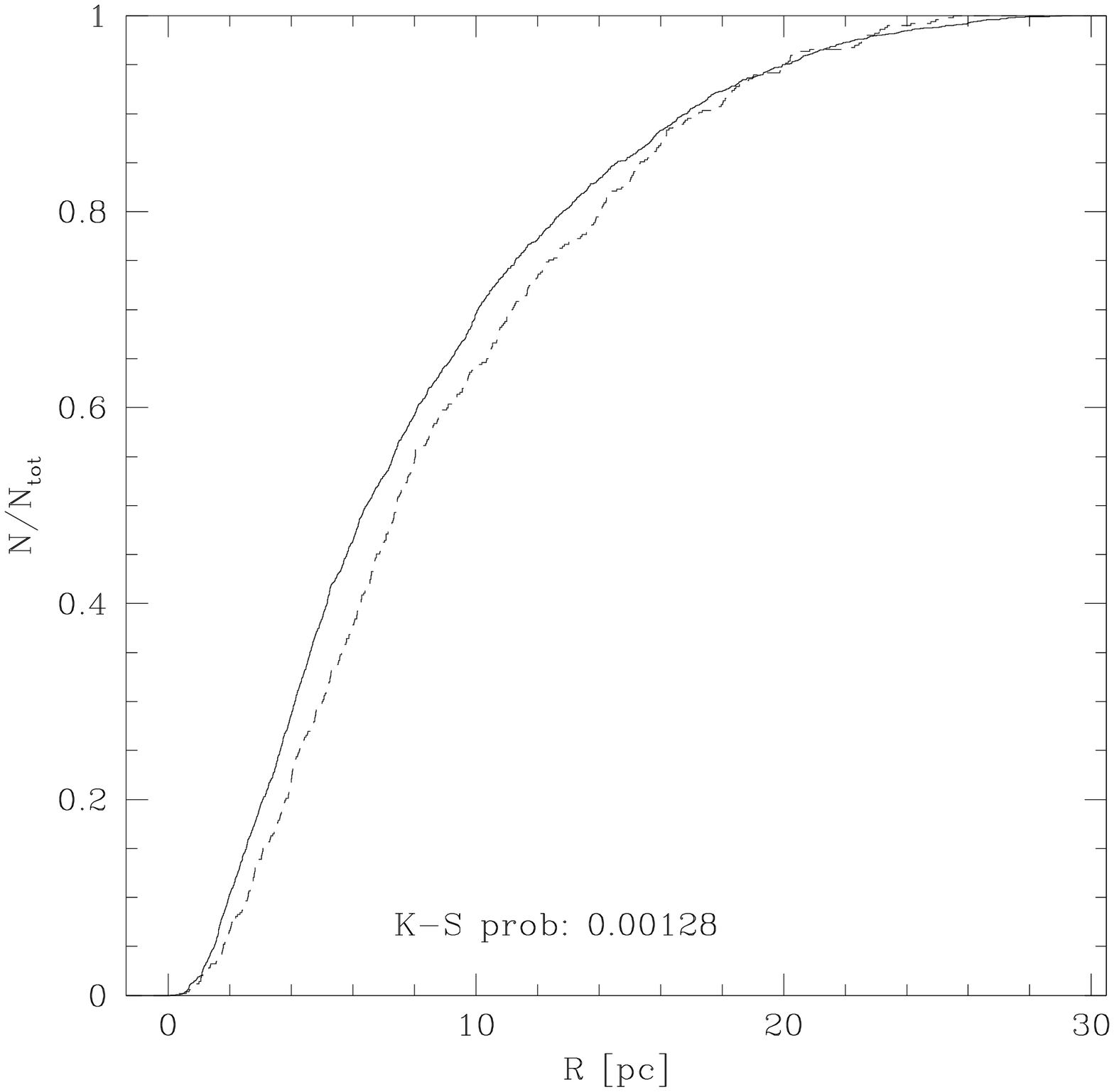}
    \caption{Cumulative three-dimensional radial distributions of old (solid curve) and young (dashed curve) WD populations
      at $\sim 12$ Gyr in models without WD birth kicks (top panel), and with kicks of $6\,{\rm km} \, {\rm s}^{-1}$ (bottom
      panel).  The young population consists of those WDs with 
      $L \geq 1.2 \times 10^{-4}\, L_\sun$ 
      (corresponding to ages $\leq 4 \, {\rm Gyr}$ for a $0.5 \, M_\sun$ WD), 
      while the old population consists of WDs with 
      $1.9 \times 10^{-5} \leq L \leq 7.9 \times 10^{-5} L_\sun$ 
      (corresponding to ages between $6$ and $10 \, {\rm Gyr}$ for a $0.5 \, M_\sun$ WD).
      The probability for obtaining the two-sided K-S statistic for each pair of distributions
      is also shown.  In the no kick case the young population is statistically significantly
      more centrally concentrated than the old population, while in the $6\,{\rm km} \, {\rm s}^{-1}$
      kick case it has a more extended radial distribution.
      \label{fig:raddist}}
  \end{center}
\end{figure}

Figure~\ref{fig:raddist} shows the cumulative three-dimensional radial distributions of WDs at $\sim 12$ Gyr 
in the models with no kick and with $6\,{\rm km} \, {\rm s}^{-1}$ kicks, broken down into ``young'' and ``old'' populations.  
The young population consists of those WDs with 
$L \geq 1.2 \times 10^{-4}\, L_\sun$ 
\citep[corresponding to ages $\leq 4 \, {\rm Gyr}$ for a $0.5 \, M_\sun$ WD;][]{1995PASP..107.1047B},
while the old population consists of WDs with 
$1.9 \times 10^{-5} \leq L \leq 7.9 \times 10^{-5} L_\sun$ 
(corresponding to ages between $6$ and $10 \, {\rm Gyr}$ for a $0.5 \, M_\sun$ WD).
In \citet{2008MNRAS.383L..20D} analogous populations are chosen with the restriction that the young WDs 
have ages less than $\sim 3$ times the local relaxation time of $0.29\,{\rm Gyr}$, so that mass segregation has had minimal
effect on their radial distribution.  The effects of mass segregation decay exponentially
in time \citep{2002ApJ...570..171F}, so they should be apparent for any division between
young and old WDs.  To maximize the evidence of the effect, we take the cut at roughly a half-mass relaxation
time, which is $\approx 5\,{\rm Gyr}$ in our models at an age of 12 Gyr.
The top panel shows the cumulative radial distribution of young and old WDs in the model
with no WD birth kicks.  The young population is statistically significantly more centrally concentrated than
the old population, as might be expected when one considers that the progenitors of the young WDs are $\sim 0.8\, M_\sun$
stars and hence more massive than the $\sim 0.5\, M_\sun$ ``old'' WDs which have had time to mass segregate
out to their larger cluster radii.
The bottom panel shows the cumulative distributions for the model with $6\,{\rm km} \, {\rm s}^{-1}$ kicks.
The young population is clearly and significantly more radially extended than the old population.
This is consistent with the results of \citet{2007MNRAS.381L..70H,2008MNRAS.385..231H}, and 
of course with the observational results of \citet{2008MNRAS.383L..20D}.

\begin{figure}
  \begin{center}
    \includegraphics[width=0.95\columnwidth]{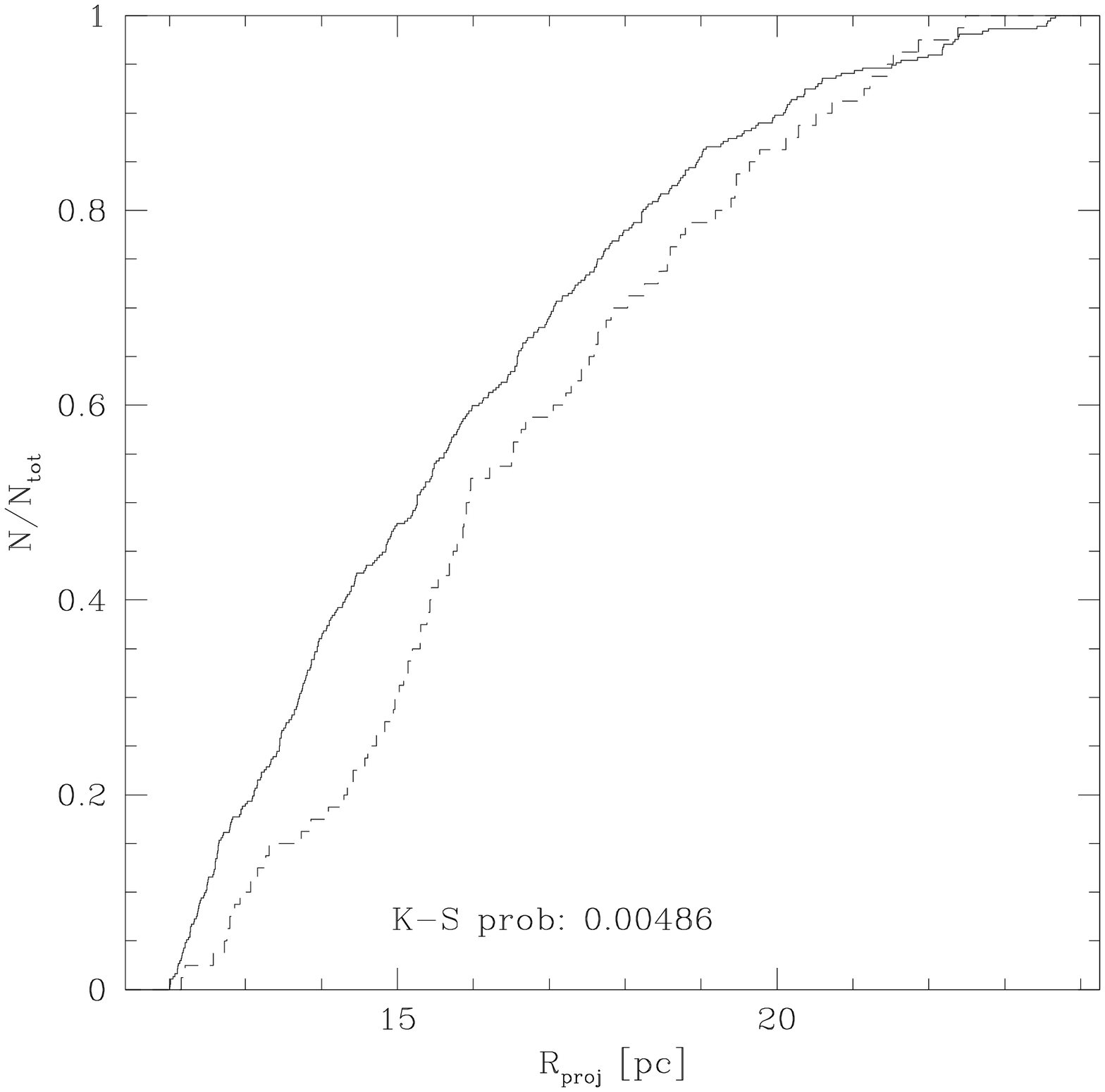}
    \caption{Cumulative projected radial distributions at $\sim 12$ Gyr between 12 and 24 pc ($1.7$ and $3.5 r_h$ in our model) 
      of old (solid curve) and young (dashed curve) WD populations
      in the model with $4 \, {\rm km}\,{\rm s}^{-1}$ WD birth kicks.  WD age cuts are as in the previous figure.
      \label{fig:rprojdist}}
  \end{center}
\end{figure}

\citet{2008MNRAS.383L..20D} observed the positions of the WDs in their sample in {\em projected} radius
between $\sim 3$ and $\sim 6\,{\rm pc}$ ($1.7$ and $3.5 r_h$).  Figure~\ref{fig:rprojdist} shows the cumulative
projected radial distributions at $\sim 12$ Gyr between 12 and 24 pc ($1.7$ and $3.5 r_h$ in our model) 
of old and young WDs in the model with $4 \, {\rm km}\,{\rm s}^{-1}$ WD birth kicks.
The young WDs are clearly more radially extended than the old population, in agreement
with what \citet{2008MNRAS.383L..20D} find in NGC 6397.  Of course, while observers
can only take snapshots of stellar positions at the current time, with numerical modeling we have
the luxury of making many statistical realizations of a model.  Figure~\ref{fig:rprojdist}
shows one of many possible realizations---some yield young WD populations that are even
more radially extended, some less.  In no case where we looked did we find one where
the old population is statistically significantly more radially extended than the young population.
We note that the average mass of the young WD population plotted in this figure is
$0.55 \, M_\sun$ and that of the old population is $0.56 \, M_\sun$, confirming that
mass segregation cannot explain the observed difference in radial distributions.
For Figure~\ref{fig:raddist} the average masses in the young and old WD populations
are $0.558$ and $0.563 \, M_\sun$, respectively, for the top panel and $0.55$ and $0.57 \, M_\sun$ for the bottom
panel.

In attempting to make quantitative conclusions about the WD kick speed, we are faced with the 
lack of detailed agreement between our model and NGC 6397.  In the cursory mapping of
parameter space we've done for this paper (consisting of $\sim 50$ simulations),
we've found that even very small kicks of 2 ${\rm km} \, {\rm s}^{-1}$ are sufficient to ``puff up'' the core
noticeably.  On the other hand, the projected radial distribution of WDs in the same
field as the observations and with the same luminosity cuts is somewhat more sensitive
to the kick speed---too low and the effect is not apparent, too high and too few
WDs remain for meaningful statistical comparisons.  Without getting mired in
detailed modeling we simply note that, as seen in Fig.~\ref{fig:raddist}, an
extended radial distribution for the young WDs compared to the older WDs is a robust consequence
of WD birth kicks.  Furthermore, a kick speed of $\sim 3$ to $5 \, {\rm km} \, {\rm s}^{-1}$,
as inferred from observations by \citet{2008MNRAS.383L..20D}, is sufficient to prolong the initial
evolutionary process of core contraction and result in a core radius at the current time
that is significantly larger than that expected from simple binary burning.

\section{Discussion}\label{sec:disc}

Through realistic numerical simulations of globular cluster evolution, we have shown
that WD birth kicks of a few ${\rm km} \, {\rm s}^{-1}$ act as an energy source that can ``puff up'' a cluster's
core potentially significantly.  In a cluster whose velocity dispersion
is comparable to the WD kick speed, $r_c/r_h$ can be increased by a factor of 
up to $\sim 10$ at late times relative to the no-kick case.  Coincidentally, theoretical
predictions of $r_c/r_h$ for a cluster whose core is supported against collapse
purely by inelastic scattering interactions of binaries (binary burning) have
recently been shown to be up to a factor of $\sim 10$ smaller than observed for
the bulk of Galactic globulars \citep[e.g.,][]{2007ApJ...658.1047F}.  WD kicks
offer a natural resolution of this large discrepancy for clusters whose
velocity dispersions are a few ${\rm km} \, {\rm s}^{-1}$, and obviate the need for alternate
energy sources whose existence remains speculative, such as IMBHs.  

Although the large core radii of some Galactic globulars
appear to be naturally explained by central IMBHs 
\citep{2006astro.ph.12040T}, there is still only circumstantial evidence for their
existence in clusters.  The evidence that exists is indirect (from cluster
velocity dispersion profiles), and for M15 and G1 can be explained equally well via standard
cluster evolution models which contain a population of neutron stars or WDs in the core
\citep{2003ApJ...582L..21B,2003ApJ...589L..25B}.  Furthermore, we should point out that 
the numerical models used to infer
the existence of central IMBHs in \citet{2006astro.ph.12040T} contain just
20,000 particles, and have IMBHs that are a fraction $M_{\rm BH}/M_{\rm clus}=0.02$
of the cluster mass.  Several independent simulations have shown that the mass of 
the stellar progenitor to the IMBH that results from a runaway collision scenario (by
far the most likely scenario for forming an IMBH in a cluster) is generically a 
fraction of $M_{\rm BH}/M_{\rm clus}=0.002$ of the total cluster mass, independent
of total cluster mass, initial central density, properties of the mass function,
and the degree of initial mass segregation 
\citep{2002ApJ...576..899P,2004ApJ...604..632G,2006MNRAS.368..121F,2006MNRAS.368..141F}.
The mass of the IMBH formed is likely to be less than the mass of its stellar 
progenitor, due to winds or other mechanisms \citep{2008A&A...477..223Y}.  Adopting
$M_{\rm BH}/M_{\rm clus}=0.002$ as an upper limit, and applying the theoretical scaling
$r_c/r_h \propto (M_{\rm BH}/M_{\rm clus})^{3/4}$
from \citet{2007PASJ...59L..11H} to the value $r_c/r_h \approx 0.3$ 
found in the simulations of \citet{2006astro.ph.12040T}, the
maximum core to half-mass radius ratio for a cluster whose core is powered by a central IMBH
should thus be $r_c/r_h \approx 0.05$.  This is appreciably smaller than the 
value $r_c/r_h \approx 0.3$ used by \citet{2006astro.ph.12040T} to infer the presence
of IMBHs, and is consistent only with the $\sim 20$\% of clusters with small $r_c/r_h$ that are classified
observationally as core-collapsed.

We have also shown that the cluster radial distribution of WDs with birth kicks is consistent
with the observations of \citet{2008MNRAS.383L..20D}.  While this certainly supports the 
interpretation of kicks as the source of the radial distribution, it is far from a direct
detection.  More detailed observations are needed to confirm the existence of the kick.
In fact, we will soon have relatively precise proper motion measurements for the bulk
of the WDs in the NGC 6397 field, which, with the help of detailed simulations,
will help pin down the properties of the WD kick.  Based on our
current understanding we can already make a few simple predictions, though.  There is no reason to
believe the magnitude of the WD kick depends on the properties of the host clusters in any 
significant way.  We thus expect two measurable consequences.  In clusters
whose velocity dispersion is significantly larger than the putative kick speed of a few ${\rm km} \, {\rm s}^{-1}$
(e.g., 47 Tuc), the radial distribution of the young WDs should follow or be more centrally concentrated than that of
the old WDs.  Observations of WD radial profiles in the high-velocity dispersion cluster Omega Cen
($\sim 20 \, {\rm km} \, {\rm s}^{-1}$ in the core) are consistent with this picture \citep{2008MmSAI..79..347C}.
In clusters whose velocity dispersion is less than a few ${\rm km} \, {\rm s}^{-1}$
(e.g., NGC 288), there should be a measurable paucity of WDs.  In fact, in the $\sim 1\,{\rm km} \, {\rm s}^{-1}$ velocity 
dispersion environments of open clusters, there is thought to be a dearth of WDs 
\citep{2002PhDT........17W}.

Finally, we note that if the kick does occur during the extended asymptotic giant branch (eAGB) 
phase, there are at least two potential complications to our modeling.  Stellar winds during the eAGB phase 
may have velocities of just $\sim 10\,{\rm km} \, {\rm s}^{-1}$ \citep{2004MNRAS.355.1348M}. 
Since the velocity dispersion is comparable to this value (or just slightly smaller), there 
is the very real possibility that the stellar wind does not escape the cluster.  For clusters
whose velocity dispersion is comparable to the WD kick speed, the ``heating'' of the cluster
via the kick is direct, and the retained wind mass should not alter our results appreciably.
However, for clusters whose velocity dispersion is less than the kick speed, the ``heating''
is indirect (via cluster mass loss, since the WDs are escaping the cluster), and the retention
of the wind mass by the cluster may appreciably reduce the ``heating'' effect.
Another potential complication is that if the mass loss during the eAGB phase takes
place on a timescale comparable to or longer than the typical stellar orbital timescale
of $\sim 10^5\,{\rm yr}$, the dynamical result is not that of a kick (as we have assumed here), 
but a weaker, adiabatic modification of the orbit.  Mass loss rates of up to $10^{-4}\,M_\sun\, {\rm yr}^{-1}$
have been detected in AGB stars (implying a mass loss timescale $\lesssim 10^4\, {\rm yr}$), but the 
general characteristics of mass loss in this phase are far from settled.

\acknowledgements

The authors thank B. Hansen and N. Ivanova for helpful discussions.
JMF acknowledges support from Chandra Postdoctoral Fellowship Award PF7-80047.
HBR acknowledges support from The Natural Sciences and Engineering Research Council of Canada,
and thanks KITP for support during a recent visit during which time this paper was conceived.
FAR acknowledges support from NASA Grant NNG06GI62G and from KITP.
This research was completed at KITP and supported in part by the NSF under Grant PHY05-51164.

\bibliographystyle{apj}
\bibliography{apj-jour,main}

\clearpage

\end{document}